\documentclass[twocolumn,showpacs,preprintnumbers,amsmath,amssymb]{revtex4}

\usepackage{graphicx}
\usepackage{dcolumn}
\usepackage{bm}

\begin{document}


\title{A GL Double Layer EM Cloaks In Broad Frequency Band And Reciprocal Law }

\author{Ganquan Xie}
 \altaffiliation[Also at ]{GL Geophysical Laboratory, USA, glhua@glgeo.com}
\author{Jianhua Li, Lee Xie, Feng Xie}%
 \email{GLGANQUAN@GLGEO.COM}
\affiliation{%
GL Geophysical Laboratory, USA
}%

\hfill\break

\date{\today}

\begin{abstract}
In this paper, we propose a novel GL double layer EM cloak in the broad frequency band. 
In short, we call it as GLWF double EM cloak.
The GLWF double layer cloak (GLWF cloak) 
consists of two sphere annular layers, two type cloak materials are proposed and installed in its each layer, respectively.  
The outer layer of the GL cloak has the invisible function in broad frequency band, 
while its inner layer has the fully absorption and rapid delay function. 
 The outer layer cloaks the Local concealment from the Global exterior EM field; 
The inner layer cloaks the Global free space region from the Local field excited inside the concealment. 
The GLWF double layer cloak overcomes the following difficulties of the single layer PS cloak.
(1) There exists no EM wavefield can be excited inside the concealment of the PS cloak, the concealment of the PS cloak is blind. Our
GL double layer cloak recovered that the EM wave field can be excited inside the concealment of the GL double layer cloak.
(2) There is exceeding light speed physical violation in PS cloak, its invisibility only in very narrow frequency band.  
Our GLWF double layer cloak corrects the violation.
(3)  The reciprocal law is satisfied in our GL double cloak media. However, the PS cloak damaged the reciprocal law..
The simulations and comparisons of the EM wave field propagation through the 
GLWF double cloak, GL double cloak and GL double layer with PS outer
layer are presented to show the advantages of the GLWF double layer EM cloak 
The GL double layer cloaks are proposed by our GL EM modeling and inversion. 
 The 3D GL EM modeling simulations for the 
double layer cloak are presented.  The copyright and patent
of the GLWF double layer cloak materials and GL EM modeling and inversion
in this paper are reserved by authors in GL Geophysical Laboratory.
\end{abstract}

\maketitle

\section{\label{sec:level1}INTRODUCTION} 
 We have proved that there exists no Maxwell wavefield can be excited by sources inside the single 
layer cloaked concealment with normal materials [1]. For covering this difficulty, we proposed a 
GL double layer cloak [2]. Its outer layer has invisibility
from the exterior light and EM wave field, never disturb the exterior field, and cloak
the Local inner layer and concealment from the Global exterior wavefield. Its inner layer cloak
 absorbs the internal wave field, such that the internal wave field, which
is excited from the sources inside the concealment, can not propagate outside of the
inner layer. Our GL double layer cloak is proposed by the GL EM modeling and inversion [3]. 
The GL double layer cloak overcomes the following difficulties of the single layer PS cloak by Pendry [4].
(1) The PS cloak damaged the EM environment of its concealment, such that there exists no EM 
wavefield can be excited inside the concealment of the PS cloak, the concealment of the PS
cloak is blind. Our GL double layer cloak recovered the normal EM environment in
its concealment, such that the EM wave field can be excited inside the concealment of the GL double layer cloak.
 (2) The two sources reciprocal law is very important principle in the electromagnetic theory and application. 
The reciprocal law is satisfied
in our GL double cloak media. However, the reciprocal law is damaged by the PS
cloak. The PS cloak is strong dispersive cloak material, in which the  
$\varepsilon _r \varepsilon _\theta   = R_3^2 (r - R_2 )^2 /r^2 /(R_3  - R_2 )^2 $, 
when $r \to R_2 $ $\varepsilon _r \varepsilon _\theta   \approx (r - R_2 )^2  \to 0.$ such that the
EM wave velocity exceed the light speed in some part of the PS cloak that is violated to the physical principle. 
The PS  cloak has invisibility only in very narrow frequency band. In the GL outer layer cloak [2] [5][6], the weak 
dispersion is obtained, when $r \to R_2 $ $\varepsilon _r \varepsilon _\theta   \approx (r - R_2 )^{1.5}  \to 0.$  In this paper, we obtain
very significant progress to overcome the difficulty. very pefect weak dispersive and degenerative rate, $\varepsilon _r \varepsilon _\theta   \approx 1/\log (r - R_2 ) \to 0,$
is obtained. Chen et al proposed an analytical method for analysis of the PS cloak [7]. 

Finding and exploration is inverse problem; Hiding and cloaking is other inverse problem.
They have close relationship.  Based on the 3D GL EM modeling simulations[3] and GL Metre Carlo inversion [5], 
We propose a novel GL double layer cloak in the broad frequency band. 

The President Professor Yuesheng Li in Sun Yat-Sen University very concerns and encourages 
our research works on the GL EM modeling method and the GL double cloak. Our paper is for 
celebrating his great scientific computational carrier in 60 years and his 80 birthday. 

The description order of this paper is as follows. We have introduced single layer cloak and our
 GL double layer cloak and main content and research cue path in Section 1. In Section 2, 
we propose a novel GL double layer cloak in the broad frequency band  The EM wave 
propagation simulations in our wide frequency band GL double layer cloak are presented in Section 3. 
The important reciprocal law in the cloak media is presented in Section 4. 
In Section 5, we will conclude our paper.

\section{\label{sec:level1}THE GL DOUBLE LAYER CLOAK MATERIALS IN  BROAD FREQUENCY BAND}
For overcoming the weakness of the single layer cloak, we proposed the GL double layer cloak in broad frequency band, in short it is called GLWF, which consists of the inner layer cloak
and outer layer cloak

\subsection {GLWF Inner layer Cloak Anisotropic Material}
On the inner sphere annular layer domain, $\Omega _{GLI}  = \left\{ {r:R_1  \le r \le R_2 } \right\} ,$  by the GL EM modeling and inversion [3][5],
we propose an anisotropic metamaterial as follows,
\begin{equation}
\begin{array}{l}
 \left[ D \right]_{GLI}  = diag\left[ {\bar \varepsilon _i ,\bar \mu _i } \right], \\ 
 \bar \varepsilon _i  = diag\left[ {\varepsilon _{r,i} ,\varepsilon _{\theta ,i} ,\varepsilon {}_{\phi ,i}} \right]\varepsilon _b,  \\ 
 \bar \mu _i  = diag\left[ {\mu _{r,i} ,\mu _{\theta ,i} ,\mu {}_{\phi ,i}} \right]\mu _b,  \\ 
 \varepsilon _{r,i}  = \mu _{r,i}  = \left( {\frac{{R_2^2  - R_1^2 }}{{R_2^2 }}} \right)\sqrt {\frac{{R_{^2 }^2  - r^2 }}{{R_2^2  - R_1^2 }}},  \\ 
 \varepsilon _{\theta ,i}  = \varepsilon _{\phi ,i}  = \mu _{\theta ,i}  = \mu _{\phi ,i}  = \sqrt {\frac{{R_2^2  - R_1^2 }}{{R_{^2 }^2  - r^2 }}} \frac{{R_2^2 }}{{R_{^2 }^2  - r^2 }}. \\ 
 \end{array}
\end{equation}
The $\Omega_{GLI}$ is called as GL inner layer cloak, the materials, $ \left[ D \right]_{GLI}  = diag\left[ {\bar \varepsilon _i ,\bar \mu _i } 
\right]$ in (1), are the anisotropic GL inner layer cloak metamaterials.
Where  the subscript $GLI$ means the GL inner layer, the symbol $diag$ denotes the diagonal matrix, $[D]_{GLI}$ is $6\times 6$ diagonal matrix,
$\bar \varepsilon _i $ is $3\times 3$ dielectric diagonal matrix in the inner layer, 
the subscript $i$ denotes the inner layer, $ \bar  \mu _i $ is $3\times 3$ magnetic 
permeability diagonal matrix in the inner layer, $ \varepsilon _{r,i} $ is the relative 
dielectric cloak metamaterial which is formulated by the fourth sub equation in (1), 
the subscript index $r,i$ denotes the dielectric is in $r$ direction and in the inner layer, 
$ \varepsilon _{\theta ,i} $ is the relative dielectric cloak metamaterial in $\theta $ 
direction and in the inner layer which is formulated by the fifth sub equation in (1), 
The $\mu _{r,i}$ is the relative permeability cloak metamaterial which is formulated 
by the fourth sub equation in (1), $\varepsilon _b$ is the basic dielectric in
free space, $\varepsilon_b  = 0.88541878176\ldots \times 10^{-11} \ \mathrm{F/m},$
$\mu _b$ is the basic magnetic permeability in
free space, $\mu_b=1.25663706 \times 10^{-6} m kg s^{-2} A^{-2},$  
other symbols in (1) have similar explanation.

The EM wavefield of sources located inside
inner layer or the cloaked concealment is completely absorbed
by the inner layer and never reaches the outside boundary
of the inner layer, also, the EM wavefield excited in concealment
is not disrupted by the cloak. The inner layer
metamaterial, in equation (1), cloaks outer space from
the local field excited in the inner layer and concealment,
which can also be useful for making a complete absorption
boundary condition to truncate infinite domain in
numerical simulation.

\subsection {GLWF Outer layer Cloak Anisotropic Material}
We proposal a new novel GLWF outer layer cloak in this section.
Let the outer sphere annular layer domain $\Omega _{GLO}  = \left\{ {r:R_2  \le r \le R_3 } \right\}$ be the GL outer layer cloak
with the following anisotropic GLWF outer layer cloak metamaterials,
\begin{equation}
\begin{array}{l}
 \left[ D \right]_{GLWFO}  = diag\left[ {\bar \varepsilon _{glwfo} ,\bar \mu _{glwfo} } \right], \\ 
 \bar \varepsilon _{glwfo}   = diag\left[ {\varepsilon _{r,,{glwfo} } ,\varepsilon _{\theta ,,{glwfo} } ,\varepsilon_{\phi ,,{glwfo} }} \right]\varepsilon _b, \\ 
 \bar \mu _{glwfo}  = diag\left[ {\mu _{r,{glwfo} } ,\mu _{\theta ,{glwfo} } ,\mu_{\phi ,{glwfo} }} \right]\mu _b,  \\ 
 \varepsilon _{r,{glwfo} } =\mu_{r ,{glwfo} } = 2\left( {r - R_2 } \right) \\
\sqrt {\log \left( {e^{1/R_3^2 } r\sqrt {\left( {R_3^2  - R_2^2 } \right)/\left( {r^2  - R_2^2 } \right)} /R_3} \right)} /R_2 /r \\ 
 \varepsilon _{\theta ,{glwfo} }  = \mu _{\theta ,{glwfo} }  = \varepsilon _{\phi ,{glwfo} }  = \mu _{\phi ,{glwfo} }  \\ 
 \approx 1/\left( {r - R_2 } \right)/  \\
\log ^{1.5} \left( {e^{1/R_3^2 } r\sqrt {\left( {R_3^2  - R_2^2 } \right)/\left( {r^2  - R_2^2 } \right)} /R_3} \right)/r \\ 
 
\end{array}
\end{equation}
The GL outer layer clok [2] is presented as follows
\begin{equation}
\begin{array}{l}
 \left[ D \right]_{GLO}  = diag\left[ {\bar \varepsilon _ {glo} ,\bar \mu _{glo}  } \right], \\ 
 \bar \varepsilon _{glo}   = diag\left[ {\varepsilon _{r,{glo} } ,\varepsilon _{\theta ,{glo} } ,\varepsilon {}_{\phi ,{glo} }} \right]\varepsilon _b, \\ 
 \bar \mu _{glo} = diag\left[ {\mu _{r,{glo} } ,\mu _{\theta ,{glo} } ,\mu {}_{\phi ,{glo} }} \right]\mu _b,  \\ 
  \varepsilon _{r,{glo} }  = \mu _{r,{glo} }  = \frac{{R_3 }}{r}\frac{{r^2  - R_2^2 }}{{r^2 }}\frac{{\sqrt {r^2  - R_2^2 } }}{{\sqrt {R_3^2  - R_2^2 } }}, \\ 
 \varepsilon _{\theta ,{glo} }  = \mu _{\theta ,{glo} }  = \varepsilon _{\phi ,{glo} }  = \mu _{\phi ,{glo} }  \\ 
  = \frac{{R_3 }}{{\sqrt {R_3^2  - R_2^2 } }}\frac{r}{{\sqrt {r^2  - R_2^2 } }}. \\ 
\end{array}
\end{equation}
We chose the PS cloak [4] as outer layer cloak
\begin{equation}
\begin{array}{l}
 \left[ D \right]_{PSO}  = diag\left[ {\bar \varepsilon _ {pso} ,\bar \mu _{pso}  } \right], \\ 
 \bar \varepsilon _{pso}   = diag\left[ {\varepsilon _{r,{pso} } ,\varepsilon _{\theta ,{pso} } ,\varepsilon {}_{\phi ,{pso} }} \right]\varepsilon _b, \\ 
 \bar \mu _{pso} = diag\left[ {\mu _{r,{pso} } ,\mu _{\theta ,{pso} } ,\mu {}_{\phi ,{pso} }} \right]\mu _b,  \\ 
\varepsilon _{r,{pso}}  = \mu _{r,{pso}}  = R_3 (r - R_2 )^2 /r^2 /(R_3  - R_2) \\ 
 \varepsilon _{\theta,{pso}}  = \mu _{\theta ,{pso}} =\varepsilon _{\phi,{pso}}  = \mu _{\phi ,{pso}} =  R_3 /(R_3 - R_2) \\ 
\end{array}
\end{equation}

Where the subscript $GLWFO$ means the $GLWF$ outer layer, $[D]_{GLWFO}$ is $6\times 6$ diagonal matrix, $\bar \varepsilon _{glwfo} $
 is $3\times 3$ dielectric diagonal matrix in the outer layer, 
the subscript $o$ denotes the outer layer, $ \bar  \mu _{glwfo}  $ is $3\times 3$ magnetic 
permeability diagonal matrix in the outer layer, $ \varepsilon _{r,{glwfo} } $ is the relative 
dielectric cloak metamaterial which is formulated by the fourth sub equation in (2), 
the subscript index $r,{glwfo} $ denotes the dielectric is in $r$ direction and in the outer layer, 
$ \varepsilon _{\phi ,{glwfo} } $ is the relative dielectric cloak metamaterial in $\phi $ 
direction and in the outer layer which is formulated by the fifth sub equation in (2), 
The $\mu _{r,{glwfo} }$ is the relative permeability cloak in outer layer and in $r$ direction which is formulated by the fourth sub equation in (2),
Similar explanation for GL outer layer cloak $GLO$ and PS cloak as outer layer $PSO$, 
The outer layer cloak provides invisibility, does not disturb
exterior EM wave field, and cloaks the local concealment from
the global exterior EM wavefield.

\subsection{GLWF Double Layer Cloak}
The GL inner cloak $\Omega_{GLI}$ domain and GL outer cloak $\Omega_{GLWFO}$ domain 
are bordering on the sphere annular surface $r=R_2$.
We assemble the $\Omega_{GLI}$ as the inner sphere annular domain and $\Omega_{GLWFO}$ as
the outer sphere annular domain and make them coupling on their interface boundary annular surface  $r=R_2$
as follows,
\begin{equation}
\begin{array}{l}
 \Omega _{GLWF}  = \Omega _{GLI} \bigcup {\Omega _{GLWFO} }  \\ 
  = \left\{ {r:R_1  \le r \le R_2 } \right\}\bigcup {\left\{ {r:R_2  \le r \le R_3 } \right\}}  \\ 
  = \left\{ {r:R_1  \le r \le R_3 } \right\}, \\ 
 \end{array}
\end{equation}
and situate the double layer anisotropic dielectric and magnetic permeability susceptibility tensors 
$\left[ D \right]_{GLWF}$ on the $\Omega _{GLWF}$  as follows,
\begin{equation}
\left[ D \right]_{GLWF}  = \left\{ {\begin{array}{*{20}c}
   {\left[ D \right]_{GLI} ,r \in \Omega _{GLI} }  \\
   {\left[ D \right]_{GLWFO} ,r \in \Omega _{GLWFO} }.  \\
\end{array}} \right.
\end{equation}
The GL cloak material $ \left[ D \right]_{GLI}  = diag\left[ {\bar \varepsilon _i,\bar \mu _i } 
\right]$ in (1) on the inner layer domain $\Omega _{GLI}$  and GL outer layer cloak material $ \left[ D \right]_{GLWFO}  = diag\left[ {\bar \varepsilon _{glwfo} ,\bar \mu _{glwfo} } 
\right]$ in (2) on outer layer domain $\Omega _{GLWFO}$  are assembled into the GL anisotropic doubled layer cloak material on
the domain $\Omega _{GLWF}$. The domain $\Omega _{GLWF}$ with the metamaterial $\left[ D \right]_{GLWF}$ in (6) 
is called as the  GL double layer cloak in broad frequency band, say $GLWF$ double cloak. Similar explanations are for $GL$ double layer
cloak coupled by $GLI$ in (1) and $GLO$ in (3), and $GLI-PSO$ double layer
cloak coupled by $GLI$ in (1) and $PSO$ in (4).

The GL cloak means that the outer layer has invisibility, never
disturb exterior wavefield and cloaks the Local concealment
from the exterior Global field, while the inner layer
cloaks outer Global space from interior wavefield excited
in the Local concealment. The properties of the double
layer material are studied by GL method simulations and
analysis in next sections.

\section{\label{sec:level1}The GL EM  Simulations and Comparison of  The GLWF Double Cloaks and Pentry Cloak}
\subsection{The Simulation Model of The GLIWFO, GLIO, GLI-PSO Double Layer Cloak}

The simulation model:
the 3D domain is $ [-0.5m,0.5 m] \times [-0.5m,0.5 m]  \times [-0.5m,0.5m]$, 
the mesh number is $201 \times 201 \times 201$, the mesh size is 0.005m. 
The electric current point source is defined as
\begin{equation}
\delta (r - r_s )\delta (t)\vec e,
\end{equation}
where the $r_s$ denotes the location of the point source,
the unit vector $\vec e$ is the polarization direction, 
the time step $dt = 0.3333 \times 10^{ - 10}$ 
second, the frequency band is from 0.05 GHz to 15 GHz, the  largest frequency $f=15 GHz$,  
the shortest wave length is $0.02m$.   The inner layer $ \Omega _{GLI} $ is
denoted by (1).
The EM GL double layer cloak $\Omega _{GLWF}  = \Omega _{GLI} \bigcup {\Omega _{GLWFO} }$  in (2),
$\Omega _{GLIO}  = \Omega _{GLI} \bigcup {\Omega _{GLO} }$  in (3), and
$\Omega _{GLI-PSO}  = \Omega _{GLI} \bigcup {\Omega _{PSO} }$  in (4)
are consist of the double spherical annular with the center in the origin and interir radius 
$R_1=0.21m$, meddle radius  $R_2=0.31m$. and exterior radius $R_3=0.45m$. The cloak is divided into 
$90 \times 180 \times 90$ cells. 
The spherical coordinate is used in the sphere 
$r \le R_3$, the Cartesian rectangular coordinate is used in outside $\Omega _{GL}$  to mesh the domain. 

\begin{figure}[b]
\centering
\includegraphics[width=0.86\linewidth,draft=false]{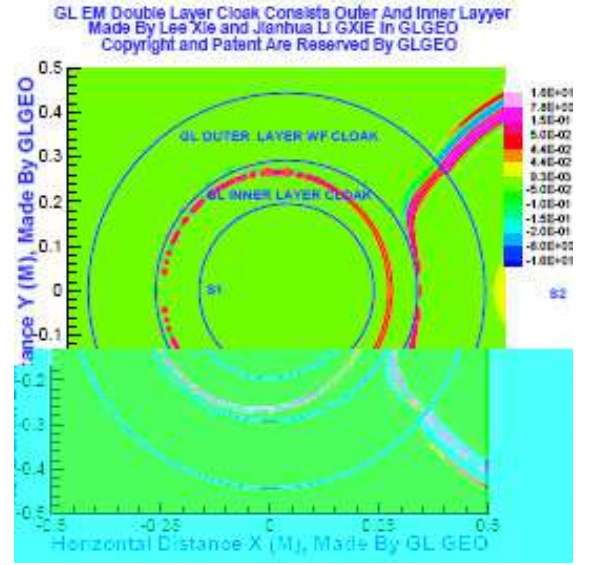}
\caption{ (color online) 
At time step $48dt$, 
front of $ {\it Second \  EM\  wave} $, $E_{xx,2}$ inside of the outer layer of GLWF cloak $R_2 \le r \le R_3$
propagates no faster than light speed, it is  faster than the wave speed in figure 2 and in figure 3. The wave front of the $ {\it First \  electric \  wave} $ , 
$E_{xx,1}$, propagates inside the inner layer,$R_1 \le r \le R_2$.}\label{fig1}
\end{figure}
\begin{figure}[b]
\centering
\includegraphics[width=0.86\linewidth,draft=false]{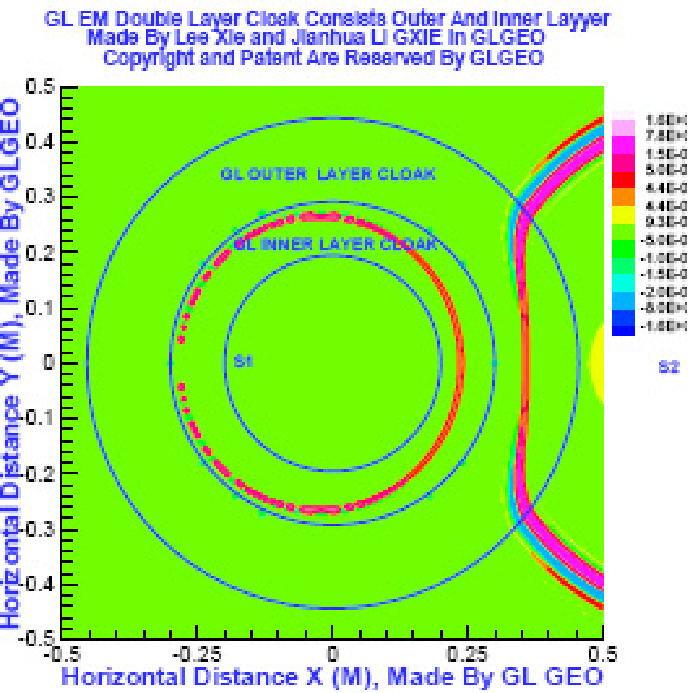}
\caption{ (color online) 
At time step $48dt$, 
front of $ {\it Second \  EM\  wave} $, $E_{xx,2}$ inside of the outer layer of GL outer layer cloak $R_2 \le r \le R_3$
propagates no faster than light speed. It is slower than $E_{xx,2}$ in figure 1 and faster than $E_{xx,2}$ in figure 3 The wave front of the
$E_{xx,1}$, propagates inside the inner layer,$R_1 \le r \le R_2$.}\label{fig2}
\end{figure}
\begin{figure}[b]
\centering
\includegraphics[width=0.86\linewidth,draft=false]{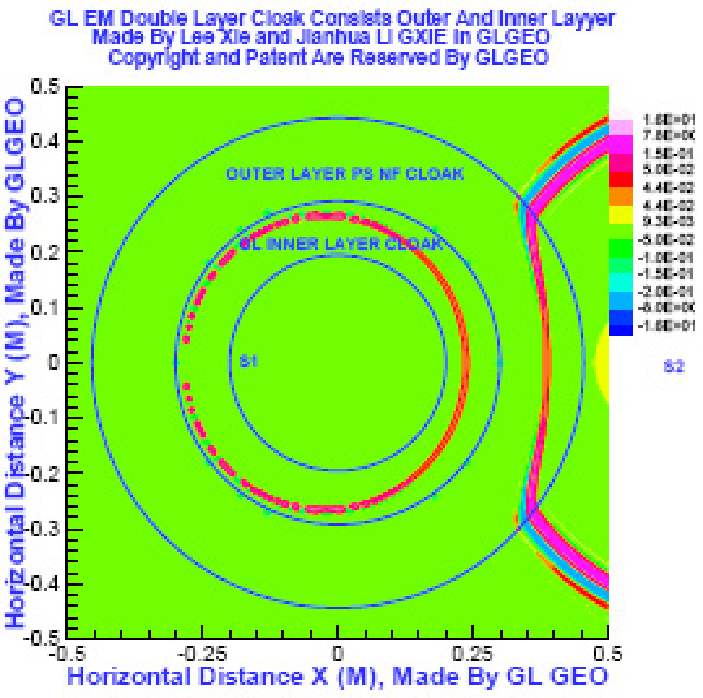}
\caption{ (color online) 
At time step $48dt$, 
front of $ {\it Second \  EM\  wave} $, $E_{xx,2}$ inside of the outer layer of PS outer layer cloak $R_2 \le r \le R_3$
propagates no faster than light speed. It is slower than $E_{xx,2}$ in figure 1 and figure 2. The wave front of the
$E_{xx,1}$, propagates inside the inner layer,$R_1 \le r \le R_2$.}\label{fig3}
\end{figure}
\begin{figure}[h]
\centering
\includegraphics[width=0.86\linewidth,draft=false]{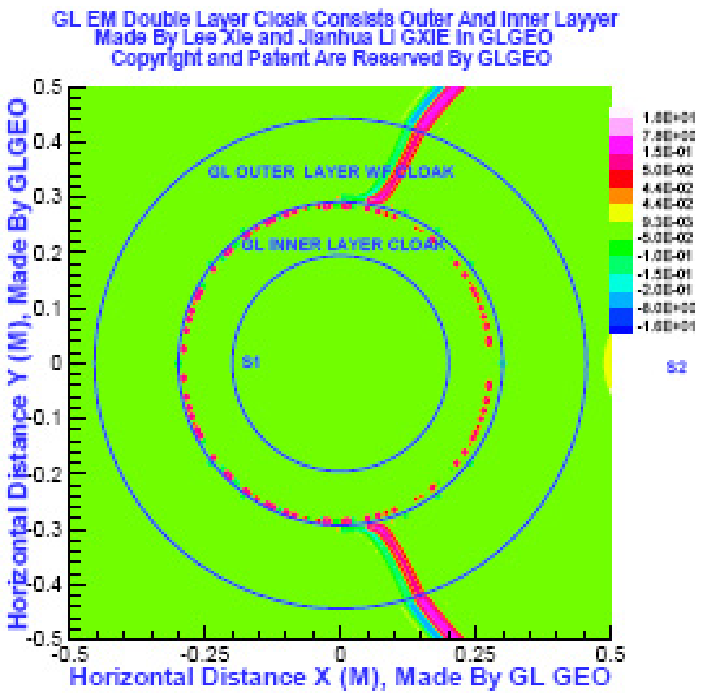}
\caption{ (color online)  At time step $74dt$, 
front of $E_{xx,2}$ inside of  GLWF outer layer cloak $R_2 \le r \le R_3$
propagates no faster than light speed, it is slower than the wave speed in figure 5 and in figure 6. The wave front of the $ {\it First \  electric \  wave} $ , 
$E_{xx,1}$, propagates inside the inner layer,$R_1 \le r \le R_2$.}\label{fig4}
\end{figure}
\begin{figure}[h]
\centering
\includegraphics[width=0.85\linewidth,draft=false]{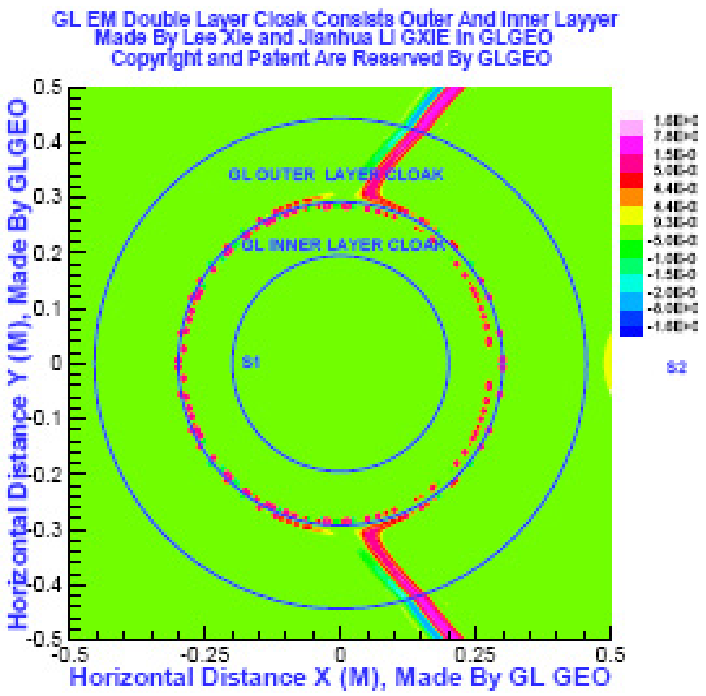}
\caption{ (color online) At time step $74dt$, 
front of  $E_{xx,2}$ inside of GL outer layer cloak $R_2 \le r \le R_3$
propagates a little faster than light speed. It is faster than $E_{xx,2}$ in GLWF in figure 4 and slower than $E_{xx,2}$ in PS in figure 6 The wave front of the
$E_{xx,1}$, propagates inside the inner layer,$R_1 \le r \le R_2$.}\label{fig5}
\end{figure}
\begin{figure}[h]
\centering
\includegraphics[width=0.85\linewidth,draft=false]{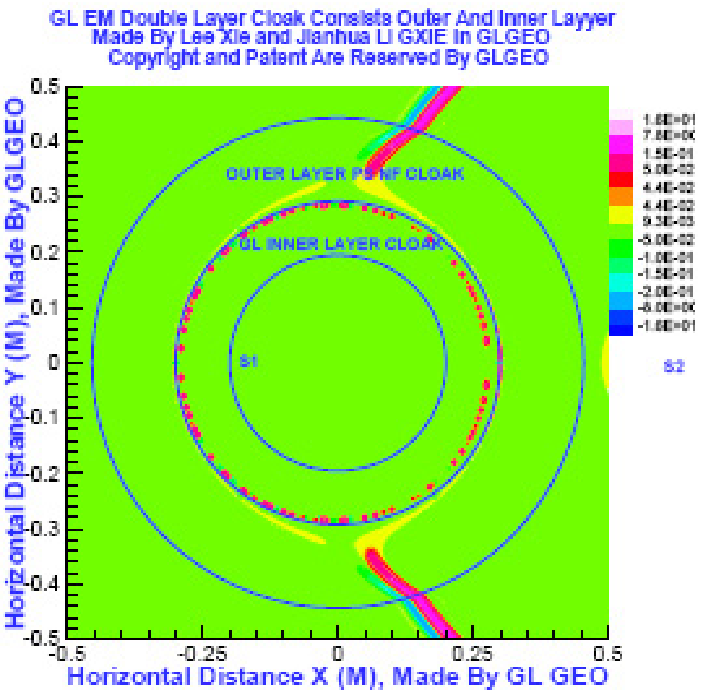}
\caption{ (color online)At time step $74dt$, 
front of $E_{xx,2}$ inside of  PS outer layer cloak $R_2 \le r \le R_3$
propagates  faster than light speed. It is faster than $E_{xx,2}$ in GLWF in figure 4 and GLO in figure 5. The wave front of the
$E_{xx,1}$, propagates inside the inner layer,$R_1 \le r \le R_2$.}\label{fig6}
\end{figure}
\begin{figure}[h]
\centering
\includegraphics[width=0.85\linewidth,draft=false]{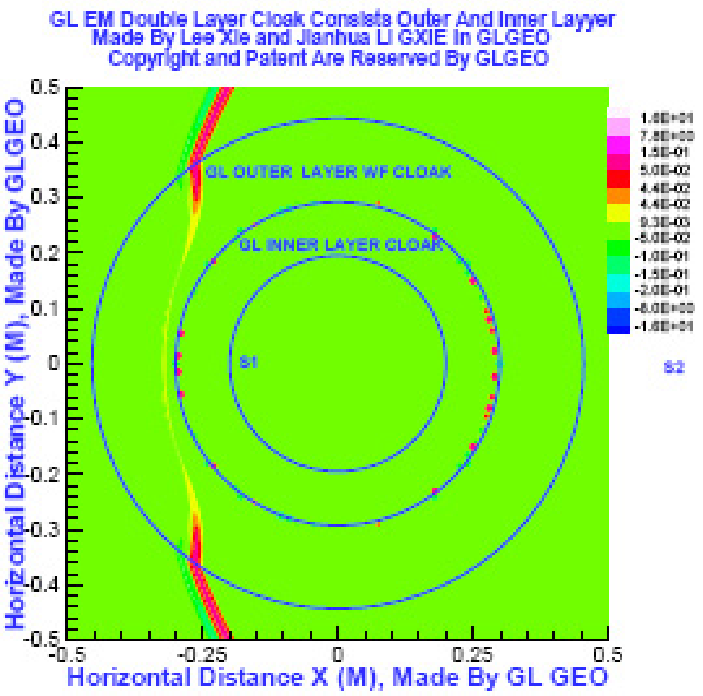}
\caption{ (color online) At time step $128dt$, 
front of $E_{xx,2}$ inside of  GLWF outer layer cloak $R_2 \le r \le R_3$
propagates slower than light speed, it is slower than the wave speed in figure 8 and in figure 9. The wave front of the $ {\it First \  electric \  wave} $ , 
$E_{xx,1}$, propagates inside the inner layer,$R_1 \le r \le R_2$.}\label{fig7}
\end{figure}
\begin{figure}[h]
\centering
\includegraphics[width=0.85\linewidth,draft=false]{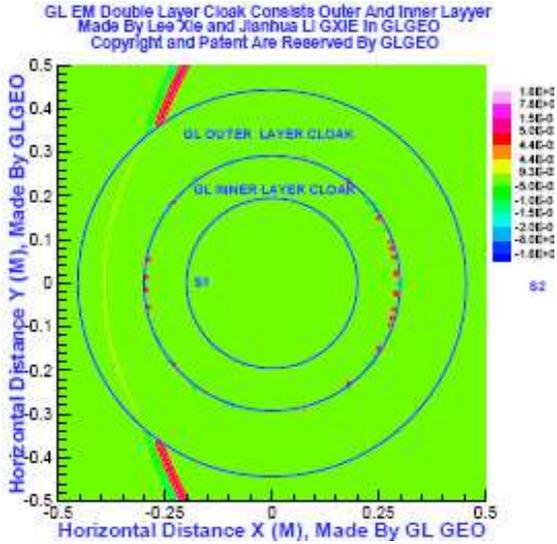}
\caption{ (color online) At time step $128dt$, 
front of  $E_{xx,2}$ inside of GL outer layer cloak $R_2 \le r \le R_3$
propagates  faster than light speed. It is faster than $E_{xx,2}$ in GLWF in figure 7 and slower than $E_{xx,2}$ in PS in figure 9 The wave front of the
$E_{xx,1}$, propagates inside the inner layer,$R_1 \le r \le R_2$.}\label{fig8}
\end{figure}
\begin{figure}[h]
\centering
\includegraphics[width=0.85\linewidth,draft=false]{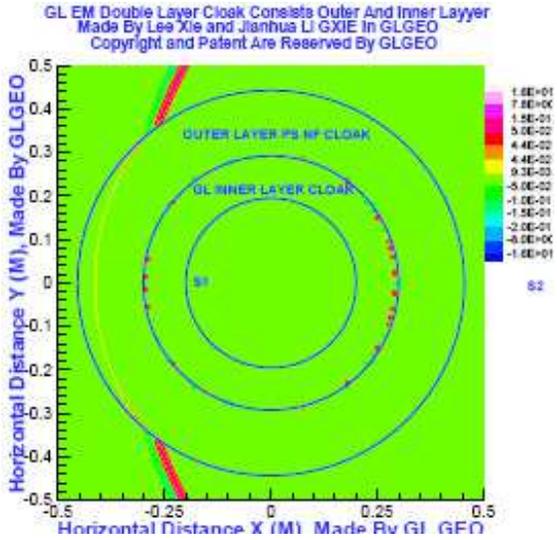}
\caption{ (color online) At time step $128dt$, 
front of $E_{xx,2}$ inside of  PS outer layer cloak $R_2 \le r \le R_3$
propagates  more faster than light speed. It is faster than $E_{xx,2}$ in GLWF in figure 7  and GLO in figure 8. The wave front of the
$E_{xx,1}$, propagates inside the inner layer,$R_1 \le r \le R_2$.}\label{fig9}
\end{figure}
\begin{figure}[h]
\centering
\includegraphics[width=0.85\linewidth,draft=false]{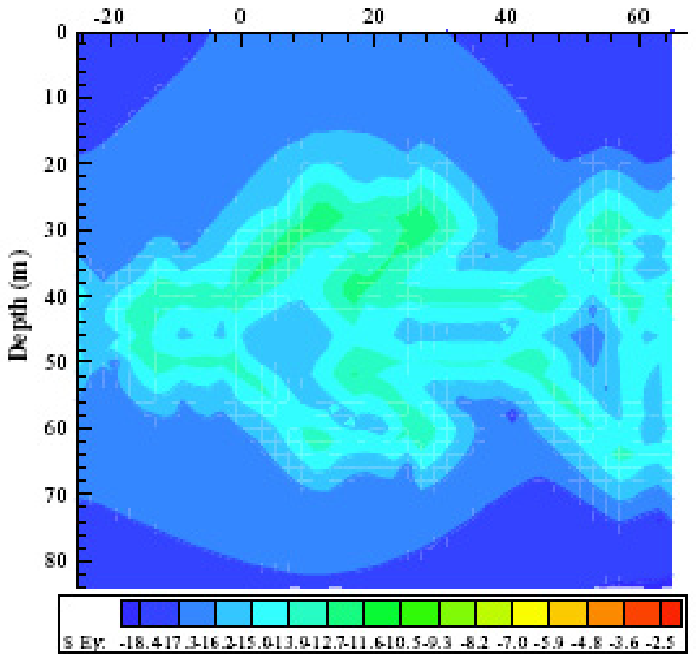}
\caption{ (color online)  
A double cloth anti detection is around the fly model, The figure 5 is from the figure 11 in paper [8] in 2001}\label{fig10}
\end{figure}
\begin{figure}[h]
\centering
\includegraphics[width=0.85\linewidth,draft=false]{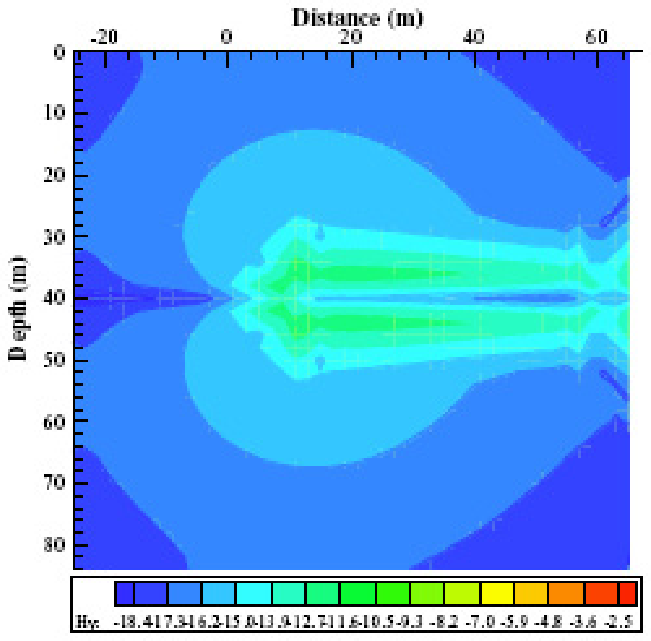}
\caption{ (color online) 
A double cloth anti detection is around the bar model, The figure 6 is from the figure 2 in paper [8] in 2001}\label{fig11}
\end{figure}
\subsection{ Point Source $S_1$ In The Concealment And Other Point Source $S_2$
In The Free Space}
The two point sources $S_1$  and $S_2$ are used to excite
the EM wave propagation through the GLWF, GLO,and GLI-PSO double layer cloaks.
The first point current source $S_1$  is located inside of the center sphere  concealment 
at $(-0.18m,0.0,0.0)$,
 the excited EM wave excied by $S_1$ is named as 
$ {\it First \  electric \  wave} $,  $E_{xx,1}$ 
The second current point source $S_2$ is located in free space at $(0.73m,0.0,0.0)$ where 
is located the right side outside of the whole GL double layer cloaks. 
The EM wave excited by the $S_2$ is named as $ {\it Second \  electric \  wave} $.$E_{xx,2}$, 
\subsection{Comparison Between EM Wave Propagation Through The GLWF, GLO,And GLI-PSO  Double Layer Cloaks}
The GL modeling simulations of the EM wave
excited by above two point sources $S_1$ and $S_2$
propagation through the GLWF, GLO,and GLI-PSO  double layer cloaks
are presented in the Figures 1-3  at $48th$ time step , in Figures 4-6 at $74th$ time step.
in Figures 7-9 at $128th$ time step, respectively.
For comparison, we arrange Figures 1-3,  Figures 4-6 , and Figures 7-9
as four figure group rows.
At time step $48dt$, 
front of  electric wave $E_{xx,2}$ inside of the outer layer of GLWF, GLO outer layer cloak, and PS outer layer, $R_2 \le r \le R_3$
are presented in Figures 1 -3 respectively. In Figure 1, $E_{xx,2}$ inside of the outer layer of GLWF, $R_2 \le r \le R_3$,
propagates no faster than light speed, it is  faster than the wave speed in figure 2 and in figure 3.  In Figure 2
electric wave  $E_{xx,2}$ inside of the outer layer of GL outer layer cloak $R_2 \le r \le R_3$
propagates no faster than light speed. It is slower than $E_{xx,2}$ in figure 1 and faster than $E_{xx,2}$ in figure 3. In Figure 3, 
front of $E_{xx,2}$ inside of the outer layer of PS outer layer cloak,  $R_2 \le r \le R_3$,
propagates no faster than light speed. It is slower than $E_{xx,2}$ in figure 1 and figure 2. The wave front of the
$E_{xx,1}$, propagates inside the inner layer,$R_1 \le r \le R_2$.
At time step $74dt$, 
front of  electric wave $E_{xx,2}$ inside of the outer layer of GLWF, GLO outer layer cloak, and PS outer layer, $R_2 \le r \le R_3$
are presented in Figures 4 -6 respectively. In Figure 4,
front of $E_{xx,2}$ inside of  GLWF outer layer cloak $R_2 \le r \le R_3$
propagates no faster than light speed, it is slower than the wave speed in figure 5 and in figure 6. In Figure 5,
front of  $E_{xx,2}$ inside of GL outer layer cloak $R_2 \le r \le R_3$
propagates little faster than light speed. It is faster than $E_{xx,2}$ in GLWF in figure 4 and slower than $E_{xx,2}$ in PS in figure 6. In Figure 6,
front of $E_{xx,2}$ inside of  PS outer layer cloak $R_2 \le r \le R_3$
propagates faster than light speed. It is faster than $E_{xx,2}$ in GLWF in figure 4 and GLO in figure 5. 
At time step $128dt$, front of  electric wave $E_{xx,2}$ inside of the outer layer of GLWF, GLO outer layer cloak, and PS outer layer, $R_2 \le r \le R_3$
are presented in Figures 7 -9 respectively. In Figure 7,
front of $E_{xx,2}$ inside of  GLWF outer layer cloak $R_2 \le r \le R_3$
propagates slower than light speed, it is slower than the wave speed in figure 8 and in figure 9. In Figure 8, front of  $E_{xx,2}$ inside of GL outer layer cloak $R_2 \le r \le R_3$
propagates little faster than light speed. It is faster than $E_{xx,2}$ in GLWF in figure 7 and slower than $E_{xx,2}$ in PS in figure 9.
In Figure 9, front of $E_{xx,2}$ inside of  PS outer layer cloak $R_2 \le r \le R_3$
propagates  more faster than light speed. It is faster than $E_{xx,2}$ in GLWF in figure 7 and GLO in figure 8. The wave front of the
$E_{xx,1}$, propagates inside the inner layer,$R_1 \le r \le R_2$.

\section{\label{sec:level1}THEORY OF RECIPROCAL LAW OF 
THE EM WAVE FIELD THROUGH THE CLOAKS}

\subsection {Theory Of 
The EM Wave Field Through The GL Double Layer Cloaks}

We propose the theoretical analysis of the interaction between the EM wave and GL cloaks in this section.

$\textbf{Statement 1:}$  Let domain $\Omega _{GL}$ in (3) and the metamaterial $D_{GL}$ in (4) be GL double layer cloak, and
 $\varepsilon {\rm  = }\varepsilon _{\rm b},\mu  = \mu _b$
be basic permittivity and permeability, respectively, 
inside of the central sphere concealment $| \vec {r'}  | < R_1$ and outside of the GL cloak $ | \vec {r'} | >R_3$,  
we have the following statements:
(1) provide the local source is located inside 
of the concealment of GL double layer cloak, $|\vec r_s |< R_1$, the excited EM wave field inside of the concealment  
never be disturbed 
by the cloak; (2) provide the local source is located inside 
of concealment or inside of the inner layer  of the GL double layer cloak,
$|\vec r_s| < R_2,$ the EM wave field is vanished outside of the inner layer of GL cloak
and is always propagating and going to the boundary $r=R_2$ and before $r=R_2$.
(3) provide the  source is located outside 
of the GL double layer cloak,  $|\vec r_s| >R_3,$ the excited EM wave field 
propagation outside of the double layer cloak as same as in free space and never be disturbed 
by the double layer cloak; (4) provide the local  source is located outside 
of double layer cloak or located inside of the outer layer of GL cloak, $|\vec r_s| >R_2,$  
the excited EM wave field never propagate into the inner layer of GL cloak and the
concealment.
\subsection {Theory Of Reciprocal Law Of 
The EM Wave Field Through The Cloaks}
$\textbf{Statement 2:}$  (1) In the domain consist of free space, single layer cloak and 
its cloaked concealment with normal material, the two sources reciprocal law is 
damaged. (2) In the domain consist of free space, single layer cloak and its cloaked 
concealment with some special double negative refractive
index metamaterial, the two sources reciprocal law is recovered, but the cloak invisibility 
function is lose. (3) In the domain consist of free space, GL double layer cloak and its cloaked 
concealment with normal material, the two sources reciprocal law is satisfied, and the cloak invisibility function is complete and sufficient
in wide frequency band.
\subsection {There Exists No Maxwell EM Wavefield Can Be Excited By Nonzero
Local Sources Inside Of The The Single Layer Cloaked Concealment With Normal Materials}
$\textbf{Statement 3:}$ Suppose that a 3D anisotropic inhomogeneous single layer
cloak domain separates the whole 3D space into three sub domains, 
one is the single layer cloak domain $\Omega _{clk}$ with the cloak material; the second one is the cloaked 
concealment domain $\Omega _{conl}$ with normal EM materials; other one is the 
free space outside of the cloak. If the Maxwell EM wavefield excited 
by a point source or local sources outside of the concealment $\Omega _{conl}$ is vanished
inside of the concealment $\Omega _{conl}$, then there excists no Maxwell EM wave field can be excited 
by the local sources inside the cloaked concealment $\Omega _{conl}$ with normal materials.

The statement 2 is proved by the GL method in author's paper [12].

\section{History and Discussions}

\subsection {History}
A double layer cloth phenomenon to prevent the GILD inversion [6][8]
detection has been observed in paper [9] in 2001 which is published in SEG online
$http://segdl.org/journals/doc/SEGLIB-home/dci/searchDCI.jsp.$ 
The double layer cloth to cloak fly from the exterior wave GILD detection is
obvious around the fly which is shown in figure 10; the double cloth around the
bar is shown in figure 11.
We developed a novel and effective Global and Local field (GL) modeling and inversion[2] [3][5] to
study the meta materials, periodic photonic crystals and condense
physics etc. wide physical sciences.
3D GL EM modeling and inversion [3] [5] and computational mirage have been presented in PIERS 2005 
and published in
proceeding of PIERS 2005 in Hangzhou, which can be downloaded from 
$http://piers.mit.edu/piersproceedings/piers2k5Proc.php$, please see the references of [2].
We developed 3D FEM for the elastic mechanics first in China in 1972[10]
and discovered the superconvergence of the 3D cubic curve isoparameter element
first in the world [11]. The 3D isoparameter element can be used for making
arbitrary curve cloak [10].
We deeply to know the merits and drawbacks of FEM. The GL method
overcomes the drawbacks of FEM and FD methods.
The history of development of FEM and GILD and GL method has been described
in [11] and reference of [2].
The 3D and 2D GL parallel software is made and patented by GLGEO.
The GL modeling and its inversion [3][5] and GL EM quantum field modeling are suitable
to solve quantization scattering problem of the electromagnetic field in the dispersive and loss metamaterials, cloaks
and more wide anisotropic materials.

\subsection {Advantages Of The GL Method}
The GL EM modeling is fully different from FEM and FD and Born approximation methods and overcome their difficulties. There is no big matrix equation to solve in GL method.
Moreover, it does not need artificial boundary and absorption condition
to truncate the infinite domain. 
Born Approximation is a conventional method in the quantum mechanics
and solid physics However , it is one iteration only in whole domain which is
not accurate for high frequency and for high contrast materials. The GL method divides the domain as a set of small 
sub domains or sub lattices. The Global field is updated by the local field 
from the interaction between the global field  
and local subdomain materials successively. Once all subdomain
materials are scattered, the GL field solution is obtained which
is much more accurate than the Born approximation. GL method
is suitable for all frequency and high contrast materials. 
When the size of the sub domain is going to zero, the GL method is
convergent and has $O(h^2)$ if the trapezoidal integral formula is used,
moreover, is has super convergence $O(h^4)$ if the Gaussian integral formula is used[10].
Chen et al proposed an analytical method for analysis of the PS cloak [7].
The GL method has double capabilities of
the theoretical analysis and numerical simulations that
has been shown in this paper.

\section{\label{sec:level1}CONCLUSIONS}
The simulations of the EM wave propagation through the GLWF. GLO, and GLI-PSO double
layer cloaks and comparison between them show that
the GLWF and GLO double layer cloak overcomes the following difficulties of the single layer PS cloak.
(1) The PS cloak damaged the EM environment of its concealment, such that there exists no EM 
wavefield can be excited inside the concealment of the PS cloak, the concealment of the PS cloak is blind. Our GL double layer cloak recovered 
the normal EM environment in
its concealment, such that the EM wave field can be excited inside the concealment of the GL double layer cloak.
(2) The PS cloak is very strong dispersive and strong degenerative cloak material. The PS cloak has invisibility only
 in very narrow frequency band. There is exceeding light speed physical violation in 
PS cloak.  Our GLWF double layer cloak corrects the violation. (3) The two sources reciprocal law is very 
important principle in the electromagnetic theory and application. 
The reciprocal law is satisfied in our GLWF and GLO double cloak media. However, the PS cloak damaged the reciprocal law.
The following physical statements are described: (1) In the domain consist of free space, 
single layer PS cloak and its cloaked concealment with normal material, the two sources 
reciprocal law is damaged. (2) In the domain consist of free space, single layer cloak and
 its cloaked concealment with some special double negative refractive
index metamaterial, the two sources reciprocal law is recovered, but the cloak invisibility 
function is lose. (3) In the domain consist of free space, GLWF and GLO double layer cloak and its 
cloaked concealment with any material, the two sources reciprocal law is satisfied, and 
the cloak invisibility function is complete, sufficient, moreover the GLWF has all advantages of the GL double layer cloak in broad frequency band.
The GLWF and GLO double layer cloak materials and the 3D and 2D GL parallel
algorithms and software are made by authors in GL Geophysical Laboratory and are patented by GLGEO and all rights are
reserved in GLGEO.

The GL method is an effective physical simulation method.
It has double abilities of the theoretical analysis and numerical simulations to study the cloak metamaterials and wide material and 
Field scattering in physical sciences.

\begin{acknowledgments}
We wish to acknowledge the support of the GL Geophysical Laboratory and thank the GLGEO Laboratory to approve the paper
publication. Authors thank to Professor P. D. Lax for his concern and encouragements  Authors thank to Dr. Michael Oristaglio
for his encouragments
\end{acknowledgments}



\end{document}